\documentclass{iaus}
\usepackage{graphicx}

\title[Magnetic fields of our Galaxy] 
{Magnetic fields of our Galaxy on large\\ and small scales}

\author[J. L. Han]   
{JinLin Han}
\affiliation{National Astronomical Observatories, Chinese Academy of
  Sciences \break Jia-20 DaTun Road, Chaoyang District, Beijing 100012,
China \break email: hjl@bao.ac.cn}

\pubyear{2007}
\volume{242}  
\pagerange{xxx -- xxx}
\date{2007 April 15th and in revised form 2007 May 29th}
\jname{Astrophysical Masers and their Environments}
\editors{J. M. Chapman \& W. A. Baan, eds.}
\begin{document}

\maketitle

\begin{abstract}
  Magnetic fields have been observed on all scales in our Galaxy, from AU to
  kpc. With pulsar dispersion measures and rotation measures, we can directly
  measure the magnetic fields in a very large region of the Galactic
  disk. The results show that the large-scale magnetic fields are aligned
  with the spiral arms but reverse their directions many times from the
  inner-most arm (Norma) to the outer arm (Perseus). The Zeeman splitting
  measurements of masers in HII regions or star-formation regions not only
  show the structured fields inside clouds, but also have a clear pattern in
  the global Galactic distribution of all measured clouds which indicates
  the possible connection of the large-scale and small-scale magnetic
  fields.

\end{abstract}

\firstsection 
\section{Introduction}

Inside our Galaxy, the diffuse interstellar medium (ISM), molecular clouds, and
very dense cloud core or HII regions, are all permeated by magnetic
fields. Many physical processes in the ISM are related to magnetic fields. For
example, magnetic fields in the diffuse medium contribute to the hydrostatic
balance and stability of the ISM (\cite{bc90}); magnetic fields are the main
agent for transporting charged cosmic-rays(e.g. \cite{tt02,ps03}), while magnetic
fields in molecular clouds obviously play an important role in the star
formation process (\cite{hc05}). However, our knowledge of the Galactic
magnetic fields is far from complete.

It is possible that the fields in clouds are enhanced when
interstellar gas contracts to form a cloud or the cloud core, so that
the observed field strength increases with gas density (\cite{cru99}). Here,
I review the observational results of large-scale magnetic fields in our
Galaxy mainly revealed by pulsar rotation measures, which are related to the
diffuse gas and the spiral structure. Small-scale magnetic fields have been
detected by other means, such as continuum polarization surveys of the
Galactic plane. I will show that the Galactic distribution of the small scale
(AU) magnetic fields detected by maser-lines from clouds are possibly
related to the large-scale Galactic magnetic field structure. In this paper 
{\it large scale} means a scale larger than the separation of spiral arms,
i.e. a scale of 1~kpc, while {\it small scale} means a smaller scale.

There are five observational tracers for the Galactic magnetic fields:
Zeeman splitting, polarized thermal emission from the dusts in clouds,
polarization of starlight, synchrotron radio emission, and Faraday rotation of
polarized sources. Observations of Zeeman splitting of spectral lines, and of
polarized thermal emission from dust at mm, sub-mm or infrared wavelengths,
have been used to detect respectively the line-of-sight strength and the
transverse orientation of magnetic fields in molecular clouds
(e.g. \cite{cru99,ncr+03,fram03}).  Starlight polarization data show that
the local field is parallel to the Galactic plane and follows the local
spiral arms (see \cite{mf70,am89,hei96}). Since we live near the edge of the
Galactic disk, we cannot have a face-on view of the global magnetic field
structure in our Galaxy through polarized synchrotron emission, as is
possible for nearby spiral galaxies (see reviews by
\cite{bbm+96,beck05}). Polarization observations of synchrotron continuum
radiation from the Galactic disk give the transverse orientation of the
field in the emission region and some indication of its strength.
Large-angular-scale features are seen emerging from the Galactic disk, for
example, the North Polar Spur (e.g.  \cite{jfr87,dhjs97,drrf99,rfr+02}), and
the vertical filaments near the Galactic Center (\cite{hsg+92,dhr+98}).
Faraday rotation of linearly polarized radiation from pulsars and
extragalactic radio sources is a powerful probe of the diffuse magnetic
field in the Galaxy (e.g. \cite{sk80,hq94,hml+06,bhg+07}). Galactic
magnetic fields both on large scales and small scales can be revealed by
these tracers.

\section{Galactic magnetic fields on large scales}

Our Galaxy is a spiral galaxy. Spiral galaxies have large-scale magnetic
fields (\cite{beck05}). In the last decade, significant progress has been
made on revealing the large-scale magnetic fields of our Galaxy in the
Galactic central region, in the Galactic halo and the Galactic disk.

Near the Galactic Center, many new non-thermal gaseous filaments have been discovered
(\cite{lnlk04,nlk+04,yhc04}). The majority of bright non-thermal filaments
are perpendicular to the Galactic plane, indicating predominantly poloidal
fields of maybe mG strength, but some filaments are not, indicating a more
complicated field structure than just the poloidal field. LaRosa et
al. (2005)\nocite{lbs+05} detected the diffuse radio emission and argued for
a weak pervasive field of tens of $\mu$G near the Galactic Center. The new
discovery of an infrared `double helix' nebula (\cite{mud06}) reinforces the
presence of strong poloidal magnetic fields merging from the rotated
circumnuclear gas disk near the Galactic Center (\cite{ym87}). With the
development of polarimetry at mm, submm or infrared wavelengths, toroidal
fields have been observed in the molecular cloud zone near the Galactic
Center (\cite{ncr+03,cdd+03}), complimenting the poloidal fields shown by
the vertical filaments. The large RMs of radio sources near the Galactic
Center (\cite{rrs05}) may indicate toroidal field structure.

From the RM distribution in the sky, Han et al. (1997, 1999)
\nocite{hmbb97,hmq99} identified the striking antisymmetry in the inner
Galaxy with respect to the Galactic coordinates after removing the RM
``outliers'' compared to their neighborhoods. The antisymmetry should result from azimuthal magnetic fields in the Galactic halo with
reversed field directions below and above the Galactic plane. Such a field
can be produced by an `A0' mode of dynamo. The observed filaments near the
Galactic Center should result from the dipole field in this dynamo
scenario. The local vertical field component of 0.2 $\mu$G
(\cite{hq94,hmq99}) may be a part of this dipole field in the solar
vicinity. At present, we have observed another 1700 radio sources in the
Northern sky by the Effelsberg 100~m telescope (Han, Reich et al. 2007, in
preparation), and wish to do more in the Southern sky with Parkes, so that
the rotation measure (RM) sky can be described quantitatively.

\begin{figure*}[hbt]
\begin{center}
\includegraphics[angle=270,width=0.97\textwidth]{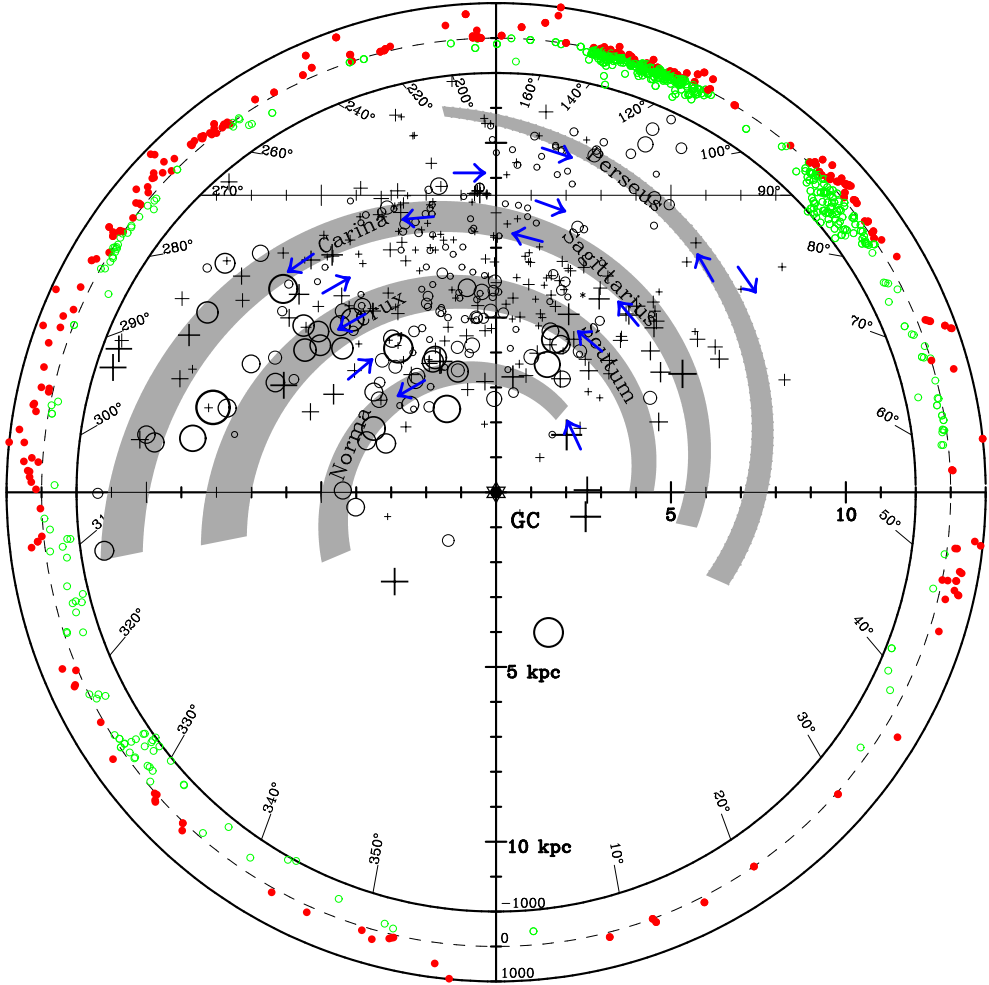}
\caption{The RM distribution of 374 pulsars with $|b|<8^{\circ}$, projected onto
the Galactic plane. The linear sizes of the symbols are proportional to the
square root of the RM values. The crosses represent positive RMs, and the
open circles represent negative RMs. The approximate locations of four
spiral arms are indicated. The large-scale structure of magnetic fields
derived from pulsar RMs are indicated by thick arrows. See Han et al. (2006)
for details. RMs of extragalactic radio sources of $|b|<8^{\circ}$ (data from
\cite{ccsk92,gdm+01,btj03,bhg+07}) are also displayed in the outskirt
ring. Positive RMs are shown by filled circles and negative RMs by open
circles. The RM limits of $\pm$1000 rad~m$^{-2}$ are set at the outer and
inner edges of the ring. As one can see from this plot, we have not many
measurements for the magnetic fields for the farther half of the Galactic
disk. The RMs of extragalactic radio sources become scarcer and scarcer in
the region of $|l|<45^{\circ}$. The fluctuations in the RM distribution
with Galactic longitude are consistent with magnetic field directions 
derived from pulsar data in the tangential regions in the 4th quadrant.}
\end{center}
\end{figure*}

Magnetic fields in a large-part of the Galactic disk have been delineated by
Faraday rotation data of pulsars, which gives a measure of the line-of-sight
component of the magnetic field. Extragalactic sources have the advantage of
large numbers but pulsars have the advantage of being spread through the
Galaxy at approximately known distances, allowing direct three-dimensional
mapping of the magnetic field. Pulsars also give a very direct estimate of
the strength of the field through normalization by the dispersion measure
(DM). The RM is defined by
$ 
\phi = {\rm RM}\; \lambda^2,
$ 
where $\phi$ is the position angle in radians of linearly polarized
radiation relative to its infinite-frequency ($\lambda = 0$) value and
$\lambda$ is its wavelength (in m). For a pulsar at distance $D$ (in pc),
the RM (in radians~m$^{-2}$) is given by
$ 
{\rm RM} = 0.810 \int_{0}^{D} n_e {\bf B} \cdot d{\bf l},
$ 
where $n_e$ is the electron density in cm$^{-3}$, ${\bf B}$ is the vector
magnetic field in $\mu$G and $d {\bf l}$ is an elemental vector along the
line of sight toward us (positive RMs correspond to fields directed toward
us) in pc. With the
$ 
{\rm DM}=\int_{0}^{D} n_e d l,
$ 
we obtain a direct estimate of the field strength weighted by the local free
electron density
\begin{equation}\label{eq_B}
\langle B_{||} \rangle  = \frac{\int_{0}^{D} n_e {\bf B} \cdot d{\bf
l} }{\int_{0}^{D} n_e d l } = 1.232 \;  \frac{\rm RM}{\rm DM}.
\label{eq-B}
\end{equation}
where RM and DM are in their usual units of rad m$^{-2}$ and cm$^{-3}$ pc
and $B_{||}$ is in $\mu$G.

Previous analysis of pulsar RM data has often used the model-fitting method
(\cite{hq94,id99}), i.e., to model magnetic field structures in all of the paths
from the pulsars to us (observer) and fit them, together with the electron density
model, to the observed RM data. {\it Significant improvement} can be obtained when both RM and DM data are available for many pulsars in a given region with similar
lines of sight. Measuring the gradient of RM with distance or DM is the most
powerful method of determining both the direction and magnitude of the
large-scale field local in that particular region of the Galaxy
(\cite{ls89,hmlq02,wck+04,hml+06}). Field strengths in the region can be
{\it directly derived} (instead of {\it modeled}) from the slope of trends
in plots of RM versus DM. Based on Equation~\ref{eq-B}, we got
\begin{equation}
\langle B_{||}\rangle_{d1-d0} = 1.232 \frac{\Delta{\rm RM}}{\Delta{\rm DM}}
\label{delta_rm_dm}
\end{equation}
where $\langle B_{||}\rangle_{d1-d0}$ is the mean line-of-sight field
component in $\mu$G for the region between distances $d0$ and $d1$,
$\Delta{\rm RM} = {\rm
RM}_{d1} - {\rm RM}_{d0}$ and $\Delta{\rm DM} ={\rm DM}_{d1} - {\rm
DM}_{d0}$.

So far, RMs of 550 pulsars have been observed
(\cite{hl87,qmlg95,hmq99,wck+04,hml+06}). Most of the new measurements are for 
the fourth and first Galactic quadrants where sources are relatively distant. This
enables us to investigate the structure of the Galactic magnetic field over a
much larger region than was previously possible. We detected
counterclockwise magnetic fields in the most inner arm, the Norma arm
(\cite{hmlq02}). A more complete analysis for the fields near the tangential
regions of most probable spiral of our Galaxy (\cite{hml+06}) gives a similar
picture for the coherent large-scale fields aligned with the spiral-arm
structure in the Galactic disk, as shown in Fig.1: magnetic fields in all
inner spiral arms are counterclockwise when viewed from the North Galactic
pole. On the other hand, at least in the local region and in the inner
Galaxy in the fourth quadrant, there is good evidence that the fields in
interarm regions are similarly coherent, but clockwise in orientation. There
are at least two or three reversals in the inner Galaxy, probably occurring
near the boundary of the spiral arms. The magnetic field in the Perseus arm
cannot be determined well. The negative RMs for distant pulsars and
extragalactic sources (see Fig.~1, also \cite{btj03}) in fact suggest the interarm fields both
between the Sagittarius and Perseus arms and beyond the Perseus arm are
predominantly clockwise.

The fluctuations in the RM distribution of extragalactic radio sources
(\cite{ccsk92,gdm+01,btj03,bhg+07}) with Galactic longitude, especially
these of the fourth Galactic quadrant, are consistent with magnetic
field directions derived from pulsar data in the tangential regions (see
Fig.~1). This implies that the dominant contribution of RMs of extragalactic
radio sources behind the Galactic disk comes from the interstellar medium
mainly in tangential regions.  However, modeling the averaged RM distribution
of scattered RM data of background radio sources behind the Galactic disk
(\cite{bhg+07}) requires fewer field reversals maybe due to the smearing effect
in the box-car averaging process over $9^{\circ}$ bins along the Galactic
longitude, larger than the separation between the inner arms.

With much more pulsar RM data now available, Han et al. (2006)
\nocite{hml+06} were able to {\it measure}, rather than {\it model}, the
regular field strength near the tangential regions in the 1st and 4th
Galactic quadrants, and then plot the dependence of regular field strength
on the Galacto-radii. Although the ``uncertainties'', which in fact reflect the
random fields, are large, the tendency is clear that fields get stronger at
smaller Galactocentric radii and weaker in interarm regions. To parameterize
the radial variation, an exponential function was used. This was chosen to give the smallest $\chi^2$ value and to avoid a singularity
at $R=0$ (for $1/R$) and unphysical values at large R (for the linear
gradient).  The function is, $ B_{\rm reg}(R) = B_0 \; \exp \left[
\frac{-(R-R_{\odot})} {R_{\rm B}} \right] , $ with the strength of the
large-scale or regular field at the Sun, $B_0=2.1\pm0.3$ $\mu$G and the
scale radius $R_{\rm B}=8.5\pm4.7$ kpc.

\section{Galactic magnetic fields on small scales}

Small-scale magnetic fields can be revealed by polarization surveys of
diffuse radio emission of the Galactic plane, and by polarization observations of
molecular clouds and supernova remnants by either linear polarization
mapping or Zeeman splitting of line emission or maser emission. The
statistics of these observations and the statistics of RMs of pulsars or
extragalactic radio sources give the overall properties of the small-scale
magnetic fields.

Polarization surveys of the Galactic plane have been comprehensively
reviewed by Reich (2007)\nocite{rei07}. See references therein.  Here I
would like to mention that the observed polarized emission of the Galactic
plane often has a scale size of about tens or hundreds of pc, and is the sum of all
contributions coming from various regions along a line of sight with
different polarization properties (i.e. polarization angle and polarization
percentage) at different distances. Emission from more distant regions
suffers from more Faraday effect produced by foreground interstellar
medium. If the emission brightness at various distances are more or less
similar, then the observed polarized emission is predominantly from local
regions. This is the case seen from observations (\cite{rei07}). Accumulating
polarized emission from different regions should ``depolarize'' each other. This is more obvious in lower frequencies. Note that disordered polarized
structures smaller than an observational beam could be smeared out by the
so-called beam depolarization. Therefore, observations with a smaller beam
would detect more polarized structures. Observations at higher
frequencies should show structures at larger distances. These polarized
structures are closely related to the magnetic field structure where the
emission is generated. The Sino-German 6~cm polarization survey
(\cite{shr+07}) of the Galactic plane using the Urumqi 25m telescope is
currently been carried out, and we have detected some magnetic structures
of Faraday screens and supernova remnants, some of which would not 
be detectable at lower frequencies (\cite{shr+07,xhr+07}).

In recent years, with development of instruments and backend technology, a
lot of molecular clouds have been directly mapped for polarized emission at
mm, submm or infrared bands, some by single dish, some by interferometers
(see review by \cite{hc05}). These maps always show field orientation on a
scale of the cloud size, i.e. a scale of pc to tens of pc. I do not have to
cover this topic as readers can find comprehensive information from the
reviews given by Crutcher (2007)\nocite{cru07} and Vlemmings
(2007)\nocite{vle07} in this volume.

Maser spots have a very small scale size of $<1$ AU.
From star formation regions or HII regions, the Zeeman splitting of the
maser lines can directly gives the field strength {\it in situ} as well as
the field direction in the line of sight. In recent years, there have been many
measurements using single dishes, e.g. by Caswell (2003)\nocite{c03}, 
or interferometers, e.g. VLBA measurements by Fish et
al. (2006)\nocite{fram03} or other measurements by Hutawarakorn \& Cohen
(2005)\nocite{hc05a}. See Han \& Zhang (2007) \nocite{hz07} for a list of
all previous measurements. It is amazing that magnetic
fields on such a small scale are possibly related to the large-scale Galactic
magnetic fields (see below).

One important tool for studying small-scale magnetic fields is to use
statistics. Detailed statistics for polarization survey data
of the Galactic plane have been rare, but would be very
useful. Statistics of the mean deviations of ``polarization vectors'' for
clouds have been used to estimate the field strength in the clouds (see
e.g. \cite{cnwk04}). Minter \&
Spangler (1996)\nocite{ms96} have worked out the structure function of a
group of RMs of extragalactic radio sources in a mid-latitude region,
showing that the magnetic fields probably follow the Kolmogorov spectrum for
a scale less than a few pc, while it becomes flatter above a few pc up to
80~pc, maybe due to the thin shape of the Galactic disk. Sun \& Han
(2003)\nocite{sh04} found that the structure function for RMs at the two
Galactic poles are flat, indicating the random RM distribution, but at lower
latitudes it becomes inclined with different slopes at different Galactic
longitudes. Haverkorn et al. (2006)\nocite{hgb+06} found that the structure
functions of RMs in the arm tangential directions have much larger slopes
than those in the interarm directions, indicating that the arm regions are
more turbulent. The outer scale is as small as only 17 pc. Han et
al. (2004)\nocite{hfm04} used statistics of pulsar RMs in a very large
region of the two third of the Galactic disk, and obtained a flat magnetic
energy spectrum for scales between 500~pc and 15~kpc, which is different
from the Kolmogorov spectrum at small-scales and should constrain the
theoretical simulations (e.g. \cite{bk05}) on how the Galactic magnetic
fields were generated and maintained.

\begin{figure}
\begin{center}
\includegraphics[width=100mm]{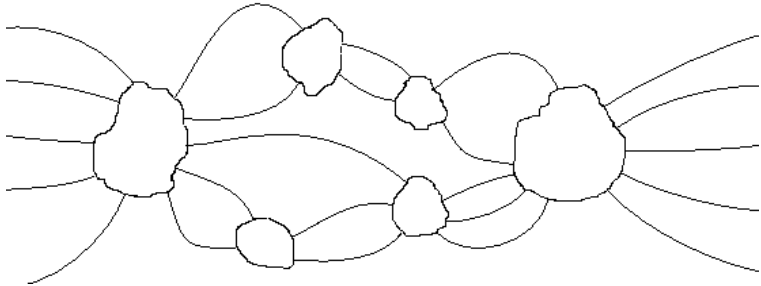}   
\caption{The magnetic fields on small scales, e.g. in clouds, are possibly
related to the large-scale fields. See also Beck et al. (1991).}
\end{center}
\end{figure}

\section{The connection of magnetic fields on large and small scales}

On one hand, molecular clouds were formed by contraction of diffuse gas in
the interstellar medium, and the magnetic fields are so enhanced that they
have the same energy as the kinetic energy (\cite{cru99}). Magnetic fields
of clouds have been observed to have an hourglass shape, which is an
indication of field direction conservation in the ISM during the contraction
(see Fig.~2 and \cite{bbb91}). On the other hand, magnetic fields in the
shell of supernova remnants are also enhanced in ISM due to supernova
explosions, as can be seen from the polarized emission of remnants. Magnetic
fields on this scale, i.e. the energy injection scale into ISM from the
kinetic, should be the strongest. On average, random magnetic fields are
stronger than the large-scale fields (\cite{rk89,hq94,hei96b,hfm04}).

However, evidence emerges for a possible relationship between the field
directions of large-scale and small-scale magnetic fields. Recently, the
observations (\cite{lgk+06}) have shown that the 
magnetic field orientations of molecular clouds seem to be preferentially parallel to the Galactic plane. From the radio observations of supernova remnants, the
bilateral remnants always tend to have the longer axis parallel to the
Galactic plane (\cite{gae98}). This could be understandable given the fact
that the Galactic magnetic fields are confined to the Galactic plane, and
the strongest component of the Galactic magnetic fields is the azimuthal
component, much stronger than the radial and vertical components
(\cite{hq94,hmq99}).

\begin{figure}[hbt]
\begin{center}
\includegraphics[angle=270,width=100mm]{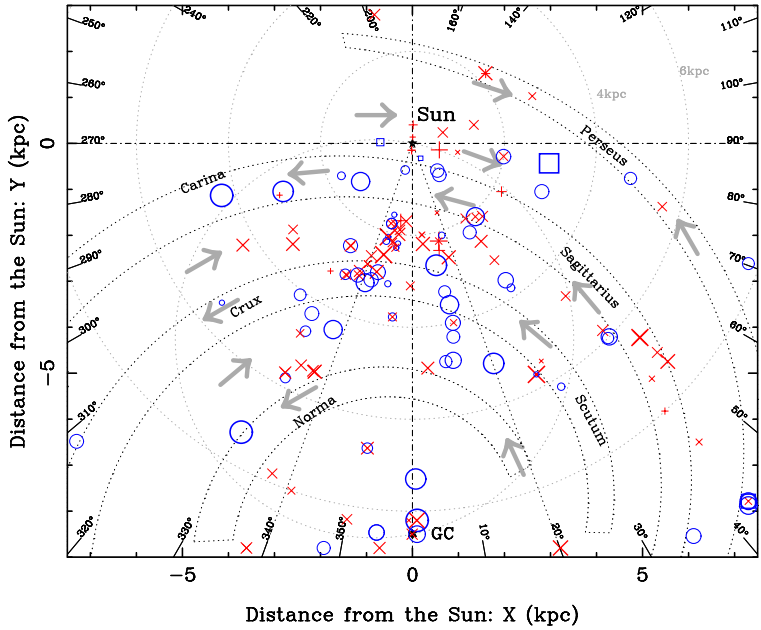}   
\caption{The medians of field measurements from Zeeman splitting of OH
masers (cross and circles) in 137 objects or HI or OH lines of 17 molecular
clouds (plus and squares) projected onto the Galactic plane, with the rough
indication of spiral arms and the magnetic field directions (arrows) derived
from pulsar RM data. The linear sizes of the symbols are proportional to the
square root of the field-strength values, with limits of 0.5 and 9~mG for
fields from the median maser fields and of 15~$\mu$G and 270~$\mu$G for
median cloud fields. The crosses or pluses on the right
($0^{\circ}<l<180^{\circ}$) represent positive $B$, i.e. the field direction going
away from observer, and circles or squares going towards us.  The symbols on
the left ($180^{\circ}<l<360^{\circ}$) are reversed, so that all crosses and
pluses are consistent with the clockwise fields viewed from the Northern
Galactic pole, and all circles and squares with counterclockwise fields. See
Han \& Zhang (2007) for details.}
\end{center}
\end{figure}

If the magnetic fields in molecular clouds are preserved from the permeated
magnetic fields in the interstellar medium during cloud formation, then the questions arise as to whether or not the magnetic fields in molecular clouds can still
``remember'' the large-scale magnetic fields in the interstellar medium? Are
they sufficiently strong that their correlation with the large-scale fields
was not destroyed by turbulence in clouds? If the answer is yes, then observations of cloud fields could be an independent approach to reveal the large-scale structure of
the Galactic magnetic fields.

Han \& Zhang (2007)\nocite{hz07} collected measurements of the magnetic
fields in molecular clouds, including Zeeman splitting data of OH masers in
clouds of HII regions and OH or HI absorption or emission lines of clouds
themselves. The data show clear structures in the sign distribution of the
median of line-of-sight components of the magnetic field (see
Fig.~3). Compared to the large-scale Galactic magnetic fields derived from
pulsar RMs, the sign distribution of the Zeeman data shows similar
large-scale field reversals. We conclude that the magnetic fields in the
clouds may still ``remember'' the directions of magnetic fields in the
Galactic ISM to some extent, and could be used as complementary tracers of
the large-scale Galactic magnetic structure.

How can such coherent magnetic field directions occur
from low density ($\sim1 {\rm cm}^{-3}$) to higher density ($\sim10^{3}{\rm cm}^{-3}$) ISM clouds, even to the highest density maser regions
($\sim10^{7}{\rm cm}^{-3}$), after a density compression of about 3, or even
10, orders of magnitude? One implication of this result is that the clouds
probably do not rotate much after they are formed as otherwise the field
directions of clouds we measured would be random. During the process of star
formation, the clouds seem to be too heavy to be rotated, although there are
jets or disks from newly formed stars which may have some dynamic
effects. Furthermore, the fields in the molecular clouds are strong enough
after the contraction that the turbulence in the clouds cannot
significantly alter the magnetic field status.

\section{Conclusive remarks}

Magnetic fields have been observed on all scales in our Galaxy. However, our
knowledge of Galactic magnetic fields and their impact on ISM physics is far from complete. Large-scale magnetic fields in some regions have been
delineated by enriched pulsar RM data (\cite{hml+06}). However, 
large-scale magnetic fields in many regions remain to be measured with more
pulsar RM data or probed by extragalactic radio sources
(\cite{bhg+07}). Small-scale fields have been measured in several ways: from
polarization surveys of the Galactic plane (tens pc or hundreds pc), 
polarization mapping of clouds and supernova remnants (pc or tens pc),
using the structure function of RMs (pc or tens pc), and Zeeman splitting
observations of line emissions (AU to pc). Although small-scale magnetic
fields appear as random or ``noise'' on the larger-scale, and are
stronger than the large-scale magnetic fields, observational evidence
already shows some physical connections between the small scale magnetic
fields and the large-scale magnetic fields.

\begin{acknowledgments}
I am very grateful to colleagues who have collaborated with me:
Dr. R.N. Manchester from Australia Telescope National Facility, CSIRO,
Prof. G.J. Qiao from Peking University (China), Prof. Andrew Lyne from
Jodrell Bank Observatory (UK), and Dr. Katia Ferri\'ere from Observatory of
Midi-Pyr\'en\'ees (France), Dr. JiangShui Zhang from GuangZhou University
(China), Dr. Willem van Straten from Swinborne University (AU). The author
thanks Dr. Jessica Chapman for careful reading of the paper. This 
work is supported by the National Natural Science Foundation of China (10521001
and 10473015).
\end{acknowledgments}

\end{document}